\documentclass[twocolumn,english,floatfix]{revtex4}

\usepackage[T1]{fontenc}
\usepackage[latin9]{inputenc}
\usepackage{float}
\usepackage{bm}
\usepackage{amsmath}
\usepackage{amssymb}
\usepackage{graphicx}
\usepackage{esint}

\begin{document}

\title{Quantum master equations for non-linear optical
 response of molecular systems}

\author{Tom\'{a}\v{s} Man\v{c}al and Franti\v{s}ek \v{S}anda}

\affiliation{Faculty of Mathematics and Physics, Charles University in Prague,
Ke Karlovu 5, CZ-121 16 Prague 2, Czech Republic}

\begin{abstract}
Generalized master equations valid for the third order response of
an optically driven multi-level electronic system are derived within Zwanzig projection formalism. 
Each of three time intervals of the response function is found to require specific
master equation and projection operator. Exact cumulant response functions 
for the harmonic profiles of the potential energy surfaces leading to Gaussian spectral diffusion are reproduced.
The proposed method accounts
for the nonequilibrium state of bath at the border of intervals, and
can be used to improve calculations of ultrafast non-linear spectra
of energy transferring systems.
\end{abstract}



\maketitle

\section{Introduction}

Non-linear response theory forms the theoretical basis for description of many important physical phenomena and experimental techniques, in particular the multi-dimensional
techniques of non-linear optical spectroscopy
\cite{MukamelBook}.
In recent years, two-dimensional (2D) coherent spectroscopy \cite{Jonas2003a},
developed first in NMR \cite{ErnstBodenhausenWokaunBook}, has been
brought into the near infra-red and optical \cite{Chobook,Hammbook}  domains. The 2D Fourier transformed
spectrum completely characterizes the third order non-linear response
of a molecular ensemble in amplitude and phase \cite{Jonas2003a}
providing thus, the maximal information accessible in a three pulse optical 
experiment. Since its first experimental realization, it has yielded
new insights into photo-induced dynamics of electronic excited states
of small molecules \cite{MIlota2009a}, polymers \cite{Collini2009a},
large photosynthetic aggregates \cite{Ginsberg2009a} and even solid
state systems \cite{Stone2009a}. For instance, long living electronic
coherence has been detected in energy transfer dynamics of photosynthetic
proteins \cite{Engel2007a}, a find that stimulated a renewed interest
in the properties of energy transfer in biological systems and its
arguably quantum nature.

Most 2D experiments are well described by the third order perturbation semiclassical
light-matter interaction response function theory.
Simulations must include  dephasing of electronic coherence  
and electronic energy transfer between levels which are both induced by interaction of the electronic
 \textit{degrees of freedom} (DOF) with environment (bath).
Response functions of model few level systems with pure dephasing
(no energy transfer between the levels) can be
calculated using second cumulant in Magnus expansion \cite{MukamelBook}, and
 expressed in terms of the \textit{energy gap correlation function}
(EGCF), $C(t)$.
For Gaussian stochastic bath, this  result is exact, and the EGCF of the electronic
transitions thus determines fully both the linear and
the non-linear (transient) absorption spectra. For energy transferring systems, such as Frenkel
excitons in photosynthetic complexes \cite{ValkunasBook}, the
exact response functions cannot be constructed by second cumulant. 
Photosynthetic complexes
are relatively large, and the proper methods to simulate finite timescale
stochastic fluctuations at finite temperatures \cite{Tanimura2006a}
carry a substantial numerical cost. Stochastic simulations can thus be readily implemented
only for small systems \cite{Ishizaki2009a,Sanda2008}. An explicite inclusion of
some important environmental modes into the Hamiltonian also increases the size of the system beyond feasibility. Practical
calculations thus require some type of reduced dynamics where only
electronic DOF are treated explicitly.
A convenient form of the reduced dynamics, master equations, is
realized by deriving kinetic equations for expectation values of
relevant quantities averaged over the DOF of the bath \cite{Rossi2002a,Axt2004a}. 
If these quantities (such as transition dipole moment) do not depend
on the bath DOF, all relevant information is carried by the {\it reduced
density matrix} (RDM).

\begin{figure} 
\includegraphics[width=1\columnwidth]{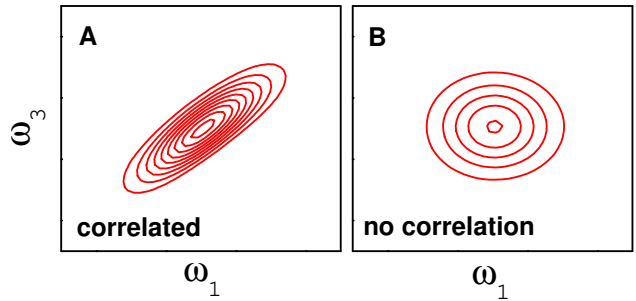}
\caption{\label{fig:tdspect} 
Two dimensional spectra of a two-level electronic system interacting with bath of harmonic oscillators at population time $T=0$, 
left panel: calculated by exact cumulant  Eq. (\ref{LSP3b}), right: by applying RDM for optical coherences (Eq. (\ref{LSP4b})).
Frequency-frequency correlation spectrum at $T=0$ defined in Ref. \cite{Nemeth2008a}
is shown for 
line-broadening function of overdamped Brownian oscillator \cite{MukamelBook} 
$g(t)=(\lambda k_B {\mathcal T} \tau_{\rm{corr}}^2 \hbar^{-1}-i\lambda \tau_{\rm{corr}})
 [e^{-t/\tau_{\rm{corr}}}-1+t/\tau_{\rm{corr}}]$
where $\lambda=100$ cm$^{-1}$, $\tau_{\rm{corr}}= 100$ fs, and the temperature is $ {\mathcal T}=300$ K. }
\end{figure}

It was demonstrated that RDM master equations
derived by projection operator technique reproduce the linear response exactly \cite{Doll2008a}. 
For higher order response, however,
the same approach neglects bath correlation between different time intervals
of the molecular photo-induced evolution separated by interaction with laser pulses (see e.g. Section  3 in Ref. \cite{Ishizaki2008a}). In other words, standard RDM master equations do not properly account for the non-equilibrium state of the phonon bath present after the second and the third laser pulse. This neglect can lead to a complete
loss of the experimentally observed dynamics in simulated 2D spectra,
such as in the case of the vibrational line shape modulation of a single electronic
transition \cite{Nemeth2008a}.
This modulation interplays with simultaneous effect of
electronic coherence. Accounting reliably for bath correlations among intervals
is therefore of utmost importance for interpretation of the role of quantum coherence in natural
light-harvesting. The difference between the 2D lineshape calculated with exact formula, and by RDM, i. e. neglecting the
correlations, is demonstrated in Fig. \ref{fig:tdspect}. We employed the definitions 2D spectra from Ref. \cite{Nemeth2008a} and the response theory from Ref. \cite{MukamelBook}.

%
%
%
%

In this Letter, we show that the exact third order response function
of a multilevel molecular system with pure dephasing can be calculated
by certain generalization of RDM master equations.  We introduce special \emph{parametric
projection operators} that enable us to derive distinct equation of
motion for each time interval of the response function. These parametric
projectors offer a systematic approach to improvement of current master equation method by
incorporating bath correlation effects into the calculation of non-linear response in energy transfer systems.

\section{Model Hamiltonian}

We model a molecular aggregate (e.g. photosynthetic antenna) as an
assembly from $N$ electronic two-level molecules coupled by resonance
interaction. Third order optical experiments on such systems address only
the collective electronic ground state $|g\rangle$ and the states
$|\bar{i}\rangle$ with one or two excited molecules of the aggregate.
Thus the $N$-molecule aggregate has $M=N+(N-1)\times N/2$ relevant excited
states. The free evolution of the system 
between the times of interaction with external light pulses is described by
Frenkel exciton Hamiltonian 
\begin{equation}
\hat{H}=\hat{Z}+\hat{V}_{g}+\varepsilon_{g}|g\rangle\langle g|+\sum_{i=1}^{M}[\varepsilon_{i}+\Delta\hat{\Phi}_{i}]|\bar{i}\rangle\langle\bar{i}|+H_{{\rm res}}.\label{eq:Hamiltonian}\end{equation}
 Here, $\hat{Z}$ represents kinetic energy of the bath DOF, $\varepsilon_{g}$ and $\varepsilon_{i}$ are the electronic energies
of the ground- and the excited 
states, respectively,  $\Delta\hat{\Phi}_{i}\equiv\hat{\Phi}_{i}-\hat{\Phi}_{g}-\langle\hat{\Phi}_{i}-\hat{\Phi}_{g}\rangle$ is the energy gap operator, and $\hat{\Phi}_{g}$
and $\hat{\Phi}_{i}$ are {\it potential energy surfaces} (PES) in
the electronic ground- and excited states, respectively. Harmonic PES represents Gaussian fluctuations of excitons. The energy gap operator
is defined so that it vanishes $\langle\Delta\hat{\Phi}_{i}\rangle\equiv tr_{B}\{\Delta\hat{\Phi}_{i}\hat{W}_{eq}\}=0$
in the ground state equilibrium $\hat{W}_{eq}=e^{-\frac{1}{k_{B}{\cal T}}\hat{H}_{B}}/z_{0}$. Here, the ground state bath Hamiltonian reads $\hat{H}_{B}=\hat{Z}+\hat{\Phi}_{g}$. The operators $\Delta \hat{\Phi}_{i}$ describe a macroscopic number of DOF so that one cannot easily treat them
explicitely in numerical calculations.
Resonance coupling 
\begin{equation}H_{{\rm res}}=\sum_{i\neq j=1}^{M}J_{ij}|\bar{i}\rangle\langle\bar{j}|
\label{eq:res_coupling}
\end{equation}
between the excited states causes electronic energy transfer between
molecules. Diagonalization of average Hamiltonian $\langle H\rangle$
is required to identify the positions of peaks in absorption spectrum,
which correspond to transition frequencies $\omega_{ig}=(\epsilon_{i}-\epsilon_{g})/\hbar$
between electronic eigenstates $|i\rangle$ with eigenenergies $\epsilon_{i}$,
$i=1,\dots,M$ and the ground state. In the very same spirit, 2D spectra
are conveniently interpreted in terms of energy transfer between these
energetic eigenstates. When resonance coupling between molecules in
the aggregate vanishes $J_{ij}\rightarrow0$ the eigenstates $|i\rangle$
coincide with excited states $|\bar{i}\rangle$, and so do the eigenenergies $\epsilon_{i}$ and energies $\varepsilon_{i}$.

\section{Linear response}

For harmonic PES and $J_{ij}\rightarrow0$ the calculation of response
functions of all orders can be carried out exactly by second order
cumulant. The linear response function 
\begin{equation}
\label{linear}
I(t)=\sum_{i}|d_{i}|^{2}\rho_{ig}(t)+c.c.
\end{equation}
is generated by time evolution of coherence elements of the RDM $\rho_{ig}(t)=e^{-i\omega_{ig}t-g_{ii}(t)}$,
which can be differentiated into the convolutionless master equation
\cite{Doll2008a,MayKuehnBook} 
\begin{equation}
\frac{\partial}{\partial t}\rho_{ig}=\left[-i\omega_{ig}-\dot{g}_{ii}(t)\right]\rho_{ig}(t).\label{convolutionless}\end{equation}
 In Eq. (\ref{linear}), $d_{i}$ are the transition dipole moments between eigenstate
$|i\rangle$ and ground state $|g\rangle$, the line broadening function
\begin{equation}
g_{ij}(t)=\frac{1}{\hbar^2}\int_{0}^{t}dt^{\prime}\int_{0}^{t^{\prime}}dt^{\prime\prime}C_{ij}(t^{\prime\prime})
\label{eq:gij}
\end{equation}
is related to the EGCF,
\begin{equation}
C_{ij}(t)=tr_{B}\{e^{(it/\hbar)\hat{H}_{B}}\Delta\hat{\Phi}_{i}e^{-(it/\hbar)\hat{H}_{B}}\Delta\hat{\Phi}_{j}\hat{W}_{eq}\},
\label{eq:Cij}
\end{equation}
in the ground state equilibrium, and $\dot{g}_{ij}\equiv({\partial g_{ij}}/{\partial t})$. Further on in this work, we assume that the electronic energy gap operators $\Delta \hat{\Phi}_{i}$ are nonzero.

\section{Master equations for non-linear response}
\begin{figure} 
\includegraphics[width=1\columnwidth]{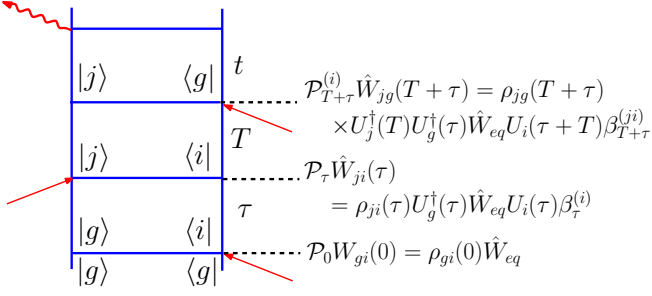}
\caption{\label{fig:Double-sided-Feynman}Double sided Feynman diagram of the
Liouville pathway $R^{ji}_{2}$ for a system with three levels $|g\rangle$,
$|i\rangle$ and $|j\rangle$. Interactions with external field are
denoted by straight arrows, the wavy arrow corresponds to emitted
field. The projectors ${\cal P}_{0}$, ${\cal P}_{\tau}$ and ${\cal P}_{T+\tau}^{(i)}$
are shown explicitly with point of their application in the response
function. }
\end{figure}

In non-linear experiment,  molecular system is probed by
three short laser pulses separated by time intervals $\tau$, $T$, and its response is monitored after time $t$.
Non-linear response function is conveniently dissected into Liouville space pathways
\cite{MukamelBook}, each representing one combination of indices of RDM during the periods $\tau$, $T$, and $t$. For our
discussion we select a pathway usually denoted as $R_{2}$ (see Fig.
\ref{fig:Double-sided-Feynman} for the corresponding double sided
Feynman diagram), i.e. the pathway which is responsible for photon
echo signal, and which contains excited state evolution during the
second interval (waiting time) $T$ \cite{MukamelBook}.  
The $R_{2}^{(ji)}$ response function involving excited states
$|i\rangle$ and $|j\rangle$ reads 
$$R_{2}^{(ji)}(t,T,\tau)=|d_{i}|^{2}|d_{j}|^{2}$$
\begin{equation}
\times tr_{B}\{{\cal U}_{jg}(t){\cal U}_{ji}(T){\cal U}_{gi}(\tau)W_{{\rm eq}}\}.\label{eq:R2_def}
\end{equation}
 Here, ${\cal U}_{ji}(t)\equiv{\cal U}_{jiji}(t)$ are abbreviated
matrix elements of the evolution superoperator ${\cal U}(t)$
which is the Green's function solution of the Liouville-von Neumann
equation with Hamiltonian, Eq. (\ref{eq:Hamiltonian}). Between each
two consecutive Green's functions in the trace of Eq. (\ref{eq:R2_def}), the action of
the external light causes transition between electronic levels $|g\rangle$,
$|i\rangle$ and $|j\rangle$. For a harmonic PES the second order
cumulant is exact, and it yields \cite{Abramavicius2009a} 
\begin{eqnarray}
R_{2}^{(ji)}(t,T,\tau) & = & |d_{i}|^{2}|d_{j}|^{2}e^{-i(\omega_{jg}t+(\omega_{jg}-\omega_{ig})T-\omega_{ig}\tau)}\nonumber \\
 &  & \times e^{-g_{ii}^{*}(\tau+T)-g_{jj}(T+t)+g_{ij}^{*}(t+T+\tau)}\nonumber \\
 &  & \times e^{-g_{ij}^{*}(t)-g_{ij}^{*}(\tau)+g_{ij}(T)}.\label{LSP3b}\end{eqnarray}
 
%
%
%
%

 By differentiating Eq. (\ref{LSP3b}) with respect to $t$ we find
the response functions to be solutions of evolution (master) equations
\begin{equation}
\frac{\partial}{\partial t}\rho_{jg}(t;T,\tau)=[-i\omega_{jg}+\mathcal{K}_{3}^{(ji)}(t,T,\tau)]\rho_{jg}(t;T,\tau)\label{eq:Eq3int}\end{equation}
 with the relaxation tensor
\begin{equation}
\mathcal{K}_{3}^{(ji)}(t,T,\tau)=-\dot{g}_{ij}^{*}(t)+\dot{g}_{ij}^{*}(t+T+\tau)-\dot{g}_{jj}(t+T).\label{eq:Tens3int}\end{equation}
and initial condition $\rho_{jg}(0;T,\tau)=R_2^{(ji)}(0,T,\tau)$.
 Master equation can be similarly found in the second interval by
differentiation of $R_{2}^{(ji)}(0,T,\tau)$ \begin{equation}
\frac{\partial}{\partial T}\rho_{ji}(T;\tau)=[-i(\omega_{jg}-\omega_{ig})+\mathcal{K}_{2}^{(ji)}(T,\tau)]\rho_{ji}(T;\tau),\label{eq:Eq2int}\end{equation}
 with relaxation tensor 
$$\mathcal{K}_{2}^{(ji)}(T,\tau)=\dot{g}_{ij}(T)-\dot{g}_{ii}^{*}(T+\tau)$$
\begin{equation}
-\dot{g}_{jj}(T)+\dot{g}_{ij}^{*}(T+\tau).\label{eq:Tens2int}
\end{equation}
and initial condition $\rho_{jg}(0;\tau)=R_2^{(ji)}(0,0,\tau)$.
In the first interval we find an analogical master equation for
$\rho_{gi}(\tau)$ which is (up to a complex conjugation) identical
to Eq. (\ref{convolutionless}). We have thus demonstrated that evolution
equations can be constructed with the structure of the convolutionless master equation, 
and that they yield exact third order response
function. Their forms, however, have now to be specific to each
interval and to each Liouville space pathway.

\section{Constructing master equations by parametric projectors}

In Ref. \cite{Doll2008a} it was pointed out that Eq. (\ref{convolutionless})
agrees with \emph{convolutionless master equation} (CLME) of Hashitsume
et al. \cite{Hashitsume1997a}, when it is evaluated to second order in $\Delta\hat{\Phi}$.
Similar microscopic derivation of Eqs. (\ref{eq:Eq3int}) to
(\ref{eq:Tens2int}) by using projection techniques would open a way
to constructing improved master equations for the electronic energy transfer
in the cases where no exact cumulant solution is known. The Zwanzig
projection technique requires to define projection superoperator by
its action on a arbitrary operator $\hat{A}$. In general form ${\cal P}\hat{A}=\sum_{ab}\hat{{\cal W}}_{ab}tr_{B}\{\langle a|\hat{A}|b\rangle\}$,
where the normalized bath matrix ($tr_{B}\{\hat{{\cal W}}_{ab}\}=1$)
can be chosen to achieve specific goals. Master equation
for the linear response, Eq. (\ref{convolutionless}), is homogenous and it corresponds
to the choice $\hat{{\cal W}}_{ab}\equiv\hat{W}_{eq}$, when ${\cal Q}\hat{W}(0)=0$
(${\cal Q}\equiv 1-{\cal P}$), and thus the inhomogeneous part of the
CLME vanishes \cite{Doll2008a} for uncorrelated initial state $\hat{W}(0)=\rho(0) W_{eq}$.
 In the second and third interval the choice $\hat{{\cal W}}_{ab}=\hat{W}_{eq}$
will not yield Eqs. (\ref{eq:Eq3int}) and (\ref{eq:Eq2int}), since the inhomogeneous
part of the CLME does not vanish. In fact, the inhomogeneous terms
cannot be eliminated in all three intervals simultaneously by any
single projector. This is a microscopic counterpart of the fact that
non-linear optical response function cannot be exactly evaluated by
simulating the reduced dynamics of the system by a single master equation.
Indeed, the homogeneous part of the CLME corresponds to only one of
the four (nonzero) contributions of the third order response function
$$R_{2}^{(ji)}(t,T,\tau)=tr_{B}\{{\cal U}_{jg}(t)$$
\begin{equation}
\times ({\cal P}+{\cal Q}){\cal U}_{ji}(T)({\cal P}+{\cal Q}){\cal U}_{gi}(\tau){\cal P}\hat{W}_{eq}\}.
\label{eq:R2ij}
\end{equation} 
The master equations (\ref{convolutionless}) to (\ref{eq:Tens2int})
can be derived when we allow different projection operator to be defined
for each time interval of the response function. Let the projection
operators ${\cal P}_{0}$, ${\cal P}_{\tau}$ and ${\cal P}_{T+\tau}^{(i)}$ in the respective intervals be chosen so that they
eliminate the inhomogeneous terms in the CLME master equation, i.e.
so that $(1-{\cal P}_{0})\hat{W}(0)=0$;  $(1-{\cal P}_{\tau})\hat{W}(\tau)=0$, etc (see  Fig. \ref{fig:Double-sided-Feynman}).
The corresponding projectors read as follows. The
\emph{first coherence projection superoperator} \begin{equation}
{\cal P}_{0}\hat{W}(t)=\hat{W}_{eq}tr_{B}\{\hat{W}(t)\}=\hat{W}_{eq}\hat{\rho}(t),\label{eq:P0}\end{equation}
 is the widely used Argyres-Kelly projector \cite{Argyres1964a}.
The second interval can be treated with the \emph{population projection
superoperator}\begin{equation}
{\cal P}_{\tau}\hat{W}(t)=\sum_{kl}\rho_{kl}(t){\cal U}_{gl}(\tau)\hat{W}_{eq}\beta_{\tau}^{(l)}|k\rangle\langle l|,\label{eq:Ptau}\end{equation}
which yields a single master equation for all RDM elements in the
second interval. Projector, Eq. (\ref{eq:Ptau}), to be applied to
a single RDM element as in Eq. (\ref{eq:Eq2int}), reads ${\cal P}_{\tau}^{(ji)}\hat{W}_{ji}(t)=\hat{\rho}_{ji}(t){\cal U}_{gi}(\tau)\hat{W}_{eq}\beta_{\tau}^{(i)}$.
In the third interval of the response a set of \emph{second coherence
projection superoperators} reflecting different bras ($\langle i|$)
during the second interval have to be defined as
$${\cal P}_{T+\tau}^{(i)}\hat{W}(t)=\sum_{k}\rho_{kg}(t){\cal U}_{ki}(T){\cal U}_{gi}(\tau)$$
\begin{equation}
\times\hat{W}_{eq}\beta_{T+\tau}^{(ki)}|k\rangle\langle g|.\label{eq:PtauT}
\end{equation}
The factors $\beta_{\tau}^{(k)}$ and $\beta_{T+\tau}^{(ki)}$ are
present to ensure that the superoperators have the projection property
$({\cal P}_{\tau})^{2}={\cal P}_{\tau}$ and $({\cal P}_{\tau+T}^{(i)})^{2}={\cal P}_{\tau+T}^{(i)}$.
From these conditions it follows e.g. $\beta_{\tau+T}^{(ii)}=\beta_{\tau}^{(i)}=1/tr_{B}\{{\cal U}_{gi}(\tau)\hat{W}_{eq}\}$.
The factor $\beta_{\tau}^{(i)}$ evaluates in the second cumulant
to $\beta_{\tau}^{(i)}=e^{g_{ii}^{*}(\tau)}$. 

An alternative to this formalism would be to account directly for the ${\cal Q}$ terms in Eq. (\ref{eq:R2ij}), and to use the time convolutionless operator identity \cite{Hashitsume1997a} for the derivation of the non-linear response as in Ref. \cite{Richter2010a}. The two approaches lead to the same exact results for a system of non-interacting molecules. For a coupled system, the quality of the equations of motion derived by the two approaches has to be compared in a separate study. However, the central idea of the projector approach, i.e. its choice in a form which satisfied $(1-{\cal P})\hat{W}=0$, is general, and it does not limit the present method to the three projectors suggested above. Depending on physical situation, we expect that specialized projectors can be introduced for interacting systems, possibly based on the comparison with the results of Ref. \cite{Richter2010a}.    

\section{Recovering the cumulant expressions}

Let us now demonstrate how these projection operators can be used
for calculation of the non-linear response function, Eq. (\ref{eq:R2_def}),
in terms of the reduced dynamics of the system. We can write $$R_{2}^{(ji)}(t,T,\tau)=|d_{j}|^{2}|d_{i}|^{2}$$
\begin{equation}
\times \bar{{\cal U}}_{jg}^{(i)}(t;T,\tau) \bar{{\cal U}}_{ji}(T;\tau)\bar{{\cal U}}_{gi}(\tau)\rho_{0},
\end{equation}
where the three reduced evolution superoperators denoted by $\bar{{\cal U}}$
are solutions of three different master equations following from the
projectors, Eqs. (\ref{eq:P0}) to (\ref{eq:PtauT}). To simplify
the matters, we will concentrate on the response function where $i=j$,
i.e. on $R_{2}^{(ii)}$.  Since we deal with a multi-level electronic
system with adiabatically separated states
($J_{ij}=0$), we can assume that $\bar{{\cal U}}_{ii}(T;\tau)=1$
and thus \begin{equation}
R_{2}^{(ii)}(t,T,\tau)=|d_{i}|^{4}\bar{{\cal U}}_{ig}^{(i)}(t;T,\tau)\bar{{\cal U}}_{gi}(\tau)\rho_{0}.\label{eq:R2_red_exact}\end{equation}
 Note that all $T$-dependence in Eq. (\ref{eq:R2_red_exact}) is
due to the second coherence parametric projector.

Now we construct the master equations for the first and
second coherence intervals (i.e. in intervals characterized by time  $\tau$ and $t$, respectively), and demonstrate that they lead to correct
response function. 
First, we split the total Hamiltonian into a sum of the pure bath
(denoted by \emph{B}), pure electronic (\emph{S}) and system--bath
interaction (\emph{S-B}) parts. Bath Hamiltonian has already
been defined above, the electronic Hamiltonian reads $\hat{H}_{S}=\epsilon_{g}|g\rangle\langle g|+\sum_{i}\epsilon_{i}|i\rangle\langle i|$,
and the system--bath coupling Hamiltonian is $\hat{H}_{S-B}=\hbar\sum_{i}\Delta\hat{\Phi}|i\rangle\langle i|.$
We define the Liouville superoperator ${\cal L}_{S-B}$ using the
corresponding Hamilton operator as ${\cal L}_{S-B}=\hbar^{-1}[\hat{H}_{S-B},\dots]_{-}$
and switch to the interaction picture induced by $\hat{H}_{0}=\hat{H}_{B}+\hat{H}_{S}$.
In the second order in ${\cal L}_{S-B}$,
and assuming that ${\cal Q}_{x}\hat{W}(0)=0$ for $x=0$, and $x=T+\tau$ we obtain the following quantum master equation

\begin{eqnarray}
 & \frac{\partial}{\partial t}{\cal P}_{x}W_{x}^{(I)}(t)=-i{\cal P}_{x}{\cal L}_{S-B}(t){\cal P}_{x}W_{x}^{(I)}(t)\nonumber \\
 & -\int\limits _{0}^{t}dt^{\prime}{\cal P}_{x}{\cal L}_{S-B}(t){\cal Q}_{x}{\cal L}_{S-B}(t^{\prime}){\cal P}_{x}W_{x}^{(I)}(t).\label{eq:convLess}\end{eqnarray}
 The time-dependence of the Liouvillian and the upper index $(I)$
of the density operator denote interaction picture with respect
to the Hamiltonian $\hat{H}_{0}=\hat{H}_{B}+\hat{H}_{S}.$

In the first coherence interval we use the projection superoperator
${\cal P}_{0}$. The construction of the $\Delta\hat{\Phi}_{i}$ operator
ensures that $tr_{B}\{\Delta\hat{\Phi}_{i}\hat{W}_{eq}\}=0$, and so the
first term in Eq. (\ref{eq:convLess}) contributes with zero. The
second term is only non-zero when no $\hat{H}_{S-B}$ occurs on the
right hand side of the density operator, and thus only the term ${\cal P}_{0}\hat{H}_{S-B}(t)\hat{H}_{S-B}(t^{\prime}){\cal P}_{0}\hat{W}^{(I)}(t)$
contributes. The evaluation leads to Eq. (\ref{convolutionless})
and consequently yields $\bar{{\cal U}}_{gi}(\tau)=e^{-g_{ii}^{*}(\tau)+i\omega_{ig}\tau}.$
A straightforward application of the second coherence projection superoperator
${\cal P}_{T+\tau}^{(i)}$ leads to
\begin{equation}
\frac{\partial}{\partial t}\rho_{ig}^{(I)}(t)=-I_{T+\tau}(t)\rho_{ig}^{(I)}(t)-M_{T+\tau}(t)\rho_{ig}^{(I)}(t),\label{eq:2ci}
\end{equation}
where 
\begin{equation}
I_{T+\tau}(t)=itr_{B}\{\Delta\hat{\Phi}_{i}(t){\cal U}_{ii}(T){\cal U}_{gi}(\tau)\hat{W}_{eq}\}e^{g_{ii}^{*}(\tau)},\label{eq:I}
\end{equation}
and
 \begin{eqnarray}
 & M_{T+\tau}(t)=\int\limits _{0}^{t}dt^{\prime}tr_{B}\{\Delta\hat{\Phi}(t)\Delta\hat{\Phi}(t^{\prime}){\cal U}_{ii}(T)\nonumber \\
 & \times{\cal U}_{gi}(\tau)\hat{W}_{eq}\}e^{g_{ii}^{*}(\tau)}-\int\limits _{0}^{t}dt^{\prime}I_{T+\tau}(t)I_{T+\tau}(t^{\prime}).\label{eq:M}
\end{eqnarray}
 The coefficients $I$ and $M$,  Eqs. (\ref{eq:I}) and (\ref{eq:M}), can
be recast into the second cumulant  by a technique described
in Ref. \cite{Zhang1998a}. A tedious but straightforward calculation
gives 
\begin{equation}
I_{T+\tau}(t)=\dot{g}_{ii}(t+T)-\dot{g}_{ii}(t)-\dot{g}_{ii}^{*}(t+T+\tau)+\dot{g}_{ii}^{*}(t)
\end{equation}
and 
\begin{equation}
M_{T+\tau}(t)=\dot{g}_{ii}(t).
\end{equation}
Combining the results of the
second order cumulant expansion of Eqs. (\ref{eq:I}) and (\ref{eq:M})
and switching back from the interaction picture we arrive at the master
equation Eq. (\ref{eq:Tens3int}). The solution
of the master equation can be easily obtained in form of the time
evolution superoperator $\bar{{\cal U}}_{ig}(t;T,\tau)$. Finally,
inserting $\bar{{\cal U}}_{gi}(\tau)$ and $\bar{{\cal U}}_{ig}(t;T,\tau)$
into Eq. (\ref{eq:R2_red_exact}) we obtain Eq. (\ref{LSP3b}) which
is the desired result. Calculation can be repeated for any
Liouville space pathway reproducing the results of
direct cumulant expansion.

Note that standard choice of ground state equilibrium for projector would yield response function
\begin{eqnarray}
R_{2}^{(ii)}(t,T,\tau) & = & |d_{i}|^{2}|d_{j}|^{2}e^{-i(\omega_{ig}t-\omega_{ig}\tau)}\nonumber \\
 &  & \times e^{-g_{ii}^{*}(\tau) -g_{ii}(t)}
\label{LSP4b}\end{eqnarray}
Comparison between 2D lineshapes simulated using  Eq. (\ref{LSP3b}) and Eq. (\ref{LSP4b}) is made in Figure \ref{fig:tdspect}. 

The projector technique developed above can be extended to finite resonance coupling by straightforward abandoning of the $J_{ij}\rightarrow 0$ limit. We will show elsewhere
\cite{Olsina2010b} that provided the optical coherences in the first
interval can be assumed independent of each other (secular approximation),
the projectors, Eqs. (\ref{eq:P0}) and (\ref{eq:Ptau}) can be used
to derive corresponding master equations. 
The complete response function involves a single master equation
for the first coherence interval and two master equations for the
waiting interval. In the second coherence interval of the response
function there are $N$ master equations in each Liouville pathways,
where $N$ is the number of single exciton states. When energy transfer
between system's eigenstates is induced by non-zero $J_{ij}$ couplings,
the application of the parametric projectors represents an approximation
whose quality depends on the previous energy transfer history of the
system. Nevertheless, unlike the usual master equations based on standard
projectors, the parametric projectors lead to correct limit in $J_{ij}\rightarrow0$.
The prescription $(1-{\cal P})\hat{W}(t)=0$ for construction of the
parametric projection operators is general, and can be used to construct
a proper projector in the cases where ${\cal P}_{\tau+T}$ would fail.

\section{Conclusions}

In this Letter, we have outlined a general method for calculation of 
non-linear response functions by quantum master
equations. We have tailored parametric projection
superoperators to derive master equations for each time interval of
the response function, and we demonstrated the validity of the method
by reproducing a well-know result exact for a multi-level system.
The results presented here can be extended to a master equation method for calculation of a general
multi-point correlation functions. This opens a new
research avenue with consequences in the theory of open systems, non-linear
spectroscopy and potentially other branches of physics.

\begin{acknowledgments}
This work was supported by the Czech Science Foundation (GACR) grant
205/10/0898 and by the Ministry of Education, Youth, and Sports of
the Czech Republic through grant ME899 and the research plan MSM0021620835.
\end{acknowledgments}

\bibliographystyle{prsty}

%
%

%
%
%



%
%


\end{document}